# Quantum Monte Carlo for minerals at high pressures: Phase stability, equations of state, and elasticity of silica

K. P. Driver[a,b], R. E. Cohen[b], Zhigang Wu[c], B. Militzer[d,b], P. López Ríos[e], M. D. Towler[e], R. J. Needs[e], and J. W. Wilkins[a]

[a]Department of Physics, The Ohio State University, 191 West Woodruff Avenue, Columbus, OH 43210, USA

[b]Geophysical Laboratory, Carnegie Institution of Washington, 5251 Broad Branch Road, NW, Washington DC 20015, USA

[c]Department of Physics, Colorado School of Mines, 1523 Illinois Street , CO 80401, USA

[d]Departments of Earth and Planetary Science and of Astronomy, University of California, Berkeley, 407 McCone Hall, CA 94720, USA

[e]TCM Group, Cavendish Laboratory, University of Cambridge, J. J. Thomson Avenue, Cambridge CB3 0HE, United Kingdom

**Silica is an abundant component of the Earth whose crystalline polymorphs play key roles in its structure and dynamics. As the simplest silicates, understanding pure silica is a prerequisite to understanding the rocky part of the Earth, its majority. First principle density functional theory (DFT) methods have often been used to accurately predict properties of silicates. Here, we study silica with quantum Monte Carlo (QMC), which until now was not computationally possible for such complex materials, and find that QMC overcomes the failures of DFT. QMC is a benchmark method that does not rely on density functionals, but rather**

**explicitly treats the electrons and their interactions via a stochastic solution of Schrödinger's equation. Using ground state QMC plus phonons within the quasiharmonic approximation from density functional perturbation theory, we obtain the thermal pressure and equations of state of silica phases up to Earth's core-mantle boundary. Our results provide the most well-constrained equations of state and phase boundaries available for silica. QMC indicates a transition to the most dense $\alpha PbO_2$ structure above the core-insulating D'' layer, suggesting the absence of significant free silica in the bulk lower mantle, which has been assumed but never proven. We also find an accurate shear elastic constant and its geophysically important softening with pressure.**

The silica minerals display a wide range of structures and properties with compression (1, 2), and are prototypical for the behavior of Earth materials from the surface through the crust and mantle. Dense polymorphs that could potentially be responsible for seismic anomalies observed in the lower mantle have prompted a number of theoretical simulations (3-6) to investigate high-pressure behavior of silica. Density functional theory (DFT) successfully predicts many qualitative features of the phase stability (3, 4), structural (5) and elastic (6) properties of silica, but it fails in fundamental ways, such as in predicting the correct structure at ambient conditions and/or accurate elastic stiffness (5, 7). In this work, we use the quantum Monte Carlo (QMC) method (8) to compute silica equations of state, phase stability, and elasticity, demonstrating improved accuracy and reliability over DFT and lack of free silica in the lower mantle.

**DFT method and drawbacks.** DFT is currently the standard model for computing materials properties and has been successful in thousands of studies for a wide range of materials. DFT is an exact theory, which states that ground state properties of a material can be obtained from functionals of the charge density alone. In practice, exchange-correlation functionals describing many-body electron interactions must be approximated. A common choice is the local density approximation (LDA), which uses a functional of the charge density in the uniform electron gas, calibrated via QMC (9). Alternate exchange-correlation functionals, such as generalized gradient approximations (GGAs) and hybrids are also useful, but there is no known functional which can provide exact results.

DFT functionals are often expected to only have problems with materials exhibiting exotic electronic structures, such as highly correlated materials, but silica has simple, closed shell covalent/ionic bonding, and should be ideal for DFT. LDA provides good results for properties of individual silica polymorphs, but it incorrectly predicts stishovite as the stable ground state structure rather than quartz. The GGA improves the energy difference between quartz and stishovite (7), giving the correct ground state structure, and this discovery was one of the results that popularized the GGA, and led to the widespread opinion that the GGA is more accurate than LDA. However, almost all other properties of silica, such as structures, equations of state, and elastic constants are often worse within the GGA (6) and other alternate approximations. Indeed, a drawback of DFT is that there exists no method to estimate the size of functional bias or know which will succeed at describing a given property.

**QMC method.** Explicit treatment of many-body electron interactions with QMC has shown great promise in studies of the electron gas, atoms, molecules, and simple solids (8). Here we address the question of whether QMC is ready for accurate computations of complex minerals by studying silica. In QMC the only essential inputs are the trial wave function (we use a determinant of DFT orbitals) and pseudopotentials for the core electrons. Two common QMC algorithms are variational Monte Carlo (VMC) and diffusion Monte Carlo (DMC). The VMC method is efficient and provides an upper bound on the ground state energy using the variational principle applied to a trial form of the many-body wave function, but usually cannot give the required chemical accuracy. DMC refines the many-body wave function statistically and projects out the ground state solution consistent with the nodes of the trial wave function. All QMC computations reported below are DMC results unless stated otherwise. Computed results, such as the total energy, are obtained with a statistical uncertainty that decreases as $1/\sqrt{N}$, where $N$ is the number electronic configurations sampled.

**Evolution of Silica Polymorphs Under Compression.** Silica phases may form in the Earth as part of subducted slabs (10) or due to chemical reactions with molten iron (11), and it is crucial to know the phase stability fields and thermodynamic equations of state to predict the effects of free silica. Quartz, a four-fold coordinated ambient phase, metastably transforms to a six-fold coordinated phase, called stishovite, above 6 GPa. Near 50 GPa, stishovite undergoes a ferroelastic transition to a $CaCl_2$-structured polymorph via instability in the elastic shear modulus (6). Experiments (1, 2) and computations (3, 4) have reported a further transition of the $CaCl_2$-structure to an $\alpha PbO_2$- structured polymorph at pressures near those obtained at the base of the mantle. To determine the role of high-pressure silica polymorphs in the mantle, we use QMC to accurately compute their thermodynamic and elastic properties.

**Results and Discussion**

**Equations of State.** Figure 1 shows the computed equations of state compared with experimental data for quartz (12, 13), stishovite/CaCl$_2$ (14, 15) and αPbO$_2$ (1, 2). The thermal equations of state are computed from the Helmholtz free energy, $F = U - TS$, by combining QMC for the internal energy, $U$, of the static lattice with DFT linear response in the quasiharmonic approximation for the vibrational energy contribution. Analytic equations of state, using the Vinet form, are obtained by fitting isotherms of the Helmholtz free energy, and pressure is determined from $P = -(\partial F / \partial V)_T$. The gray shading of the QMC curves indicates one standard deviation of statistical errors from the Monte Carlo data. The QMC results generally agree well with diamond-anvil-cell measurements at room temperature, as do the corresponding DFT calculations using the GGA functional of Wu and Cohen (WC) (16).

**Enthalpy Differences.** A phase transition between two phases is predicted where the difference in free energy (or enthalpy at T=0) is zero, as shown in Figure 2. Uncertainties in the differences determine how well phase boundaries are constrained. The one-sigma statistical error on the QMC enthalpy difference is 0.5 GPa for the quartz-stishovite transition and 8 GPa for the CaCl$_2$-αPbO$_2$ transition. The error on the latter is larger because the scale of the enthalpy change for the quartz-stishovite transition is about a factor of ten larger than for $CaCl_2$-$\alpha PbO_2$. In both transitions, variation in the DFT result with functional approximation is large. For the metastable quartz-stishovite transition, LDA incorrectly predicts stishovite to be the stable ground state, WC under predicts the quartz-stishovite transition pressure by 67%, and the GGA of Perdew, Burke, and Ernzerhof (PBE) (17) matches the QMC result. (The mineral coesite is stable from 2-7.5 GPa, but we did not study it due to its structural complexity, which is monoclinic, with 24 atoms per unit cell. The DFT discrepancy has centered on the metastable quartz-stishovite transition, which we consider here.) For the $CaCl_2$-$\alpha PbO_2$ transition, the same three DFT approximations agree with QMC within statistical uncertainty. Longer runs would refine the transition pressure, but the four times longer runs required to halve the error in the energies are not currently feasible; as larger computers become more common, much greater accuracy should be routinely possible. Even the present results apparently have greater reliability than different experimental determinations of this transition (Fig. 3). This may be due to the difficulty in demonstrating rigorous phase transition reversals as well as pressure and temperature gradients and uncertainties in state-of-the-art experiments.

**Improvement of DFT with QMC.** One of the goals of this work is to find what is needed to improve DFT functionals. Previous analysis of LDA and GGA charge densities of quartz and stishovite and densities of states showed no changes in the predicted bonding, but rather changes in the atomic regions, akin to slightly different predictions of effective ionic size (18). Consistent with this analysis, we find that, for each phase, QMC energies and pressures are shifted by roughly a constant relative to DFT at every volume.

For instance, QMC gives a lower energy (pressure) of 1.6 eV/$SiO_2$ (1.5 GPa) for quartz, 1.3 eV/$SiO_2$ (3.8 GPa) for stishovite and 1.2 eV/$SiO_2$ (6.2 GPa) for $\alpha PbO_2$ for any given cell volume. The constant shift suggests one could correct DFT computations for silica and potentially other systems by simply doing a single QMC calculation for each phase. Then one could use the corrected DFT for all other computations and significantly reduce computational expenditures.

**Phase Boundaries.** The QMC quartz-stishovite phase boundary shown in Figure 3a agrees well with thermodynamic modelling of shock data (19, 20) and thermocalorimetry measurements (21, 22). In addition, the QMC boundary crosses the silica melting curve (23) at the intersection of the quartz and stishovite melts, suggesting that the DFT phonons are accurate even at high temperatures. The WC boundary is about 4 GPa too low in pressure.

The QMC statistical error in the $CaCl_2$- $\alpha PbO_2$ phase boundary shown in Figure 3b is larger than for the quartz-stishovite transition because the free energy difference between the two phases is small. The WC boundary lies within the QMC statistical error. Previous DFT work also shows that the LDA boundary lies near the upper range of our QMC boundary and PBE produces a boundary 10 GPa higher than the LDA (4). The diamond-anvil-cell experiments have constrained the transition to $\alpha PbO_2$ to lie between 65 and 120 GPa near the mantle geotherm (2500 K). The only measurement of the boundary slope is negative, but QMC and WC as well as previous DFT studies (4) predict a positive slope.

**Geophysical Significance.** The QMC $CaCl_2$-$\alpha PbO_2$ boundary indicates that the transition to $\alpha PbO_2$, within a two-sigma confidence interval, occurs at a depth of 2000-2650 km (86-122 GPa) and temperature of 2300-2600 K in the lower mantle. This places the transition 50-650 km above the D″ layer, a thin boundary surrounding Earth's core ranging from a depth of ~2700 to 2900 km. The DFT boundaries all lie within the QMC two-sigma confidence interval, with PBE placing the transition most near the D″ layer. Free silica in D″, such as in dead slabs or mantle-core reaction zones would have the $\alpha PbO_2$ structure. However, based on our results, the absence of a global seismic anomaly above D″ suggests that there is little or no free silica in the bulk of the lower mantle. The $\alpha PbO_2$ phase is expected to remain the stable silica phase up to the core-mantle boundary. With respect to post-perovskite, $MgSiO_3$, at 120 GPa in D″, QMC predicts $\alpha PbO_2$ has 1-2% lower density and 6-7% larger bulk sound velocity, which may provide enough contrast to be seen seismically if present in appreciable amounts.

**Shear Modulus Softening.** Most of our information about the deep Earth comes from seismology, and we need elastic moduli to compute sound velocities in the Earth. Much work has been done using DFT to compute and predict elastic moduli for minerals in the Earth, but there is significant uncertainty in the predicted moduli because different density functionals give significantly different values. However, the energy scale for crystal strains needed to compute the shear modulus is small; meV accuracy is needed, requiring lengthy Monte Carlo statistics accumulation runs to obtain meaningful statistical error bars. We test the feasibility of using QMC for computing the softening of the shear modulus ($c_{11}$-$c_{12}$) elastic constant in stishovite, which drives the ferroelastic transition to CaCl$_2$ (6). We determine the elastic constant by computing the total energy as a function of volume-conserving strains of up to 3% strain through the relation $c_{ij} = (1/V)(\partial^2 U/\partial \varepsilon_i \partial \varepsilon_j)$. Brute force computation of the QMC total energies is very computationally expensive, requiring roughly 100-1000 times more CPU time than a corresponding DFT calculation. Correlated sampling techniques could drastically reduce the computational cost, but there is yet no implementation of these methods for elasticity.

Figure 4 shows our computations for the stishovite shear modulus, which alone used over 3 million CPU hours. For a well-optimized trial wave function, VMC often comes close to matching the results of DMC. We therefore chose to compute the shear modulus at several pressures using VMC, and checked only the endpoints with the more accurate DMC method. Both QMC and DFT results correctly describe the softening of the shear modulus, indicating the zero temperature transition to CaCl$_2$ near 50 GPa. Radial x-ray diffraction data (24) lies lower than the calculated results, however, discrepancies can arise in the experimental analysis depending on which strain model is chosen to fit the raw data. The zero pressure Brillouin data (25) agrees well with DMC.

A number of other thermodynamic parameters can be computed from the thermal equation of state. Table I summarizes ambient computed and available experimentally measured values of the Helmholtz free energy, $F_0$ (Ha/SiO$_2$), volume, $V_0$ (Bohr$^3$/SiO$_2$), bulk modulus, $K_0$, pressure derivative of the bulk modulus, $K'_0$, thermal expansivity, $\alpha$ (K$^{-1}$), heat capacity, $C_p/R$, Grüneisen ratio, $\gamma$, volume dependence of the Grüneisen ratio, $q$, and the Anderson-Grüneisen parameter, $\delta_T$, for the quartz, stishovite, and αPbO$_2$ phases of silica. QMC generally offers excellent agreement with experiment for each of these parameters.

**Conclusions**

In conclusion, we have presented QMC computations of silica equations of state, phase stability, and elasticity. Results demonstrate the improved accuracy and reliability of QMC relative to DFT and indicate there is not significant free silica in the bulk lower mantle. However, DFT currently remains the method of choice for computing material properties because of its computational efficiency, but we have shown that QMC is feasible for computing thermodynamic and elastic properties of complex minerals. DFT is generally successful, but does display failures independent of the complexity of the electronic structure and sometimes shows strong dependence on functional choice. With the current levels of computational demand and resources, one can use QMC to spot-check important DFT results to add confidence at extreme conditions or provide insight into improving the quality of density functionals. In any case, we expect QMC to become increasingly important and common as next generation computers appear and have a great impact on computational materials science.

**Methods**

**Pseudopotentials.** Pseudopotentials were used to replace the core electrons of the atoms by an effective potential to improve computational efficiency. All calculations used norm-conserving, nonlocal pseudopotentials constructed with the OPIUM (30) code and were generated with the WC functional. The silicon potential has a Ne core with 3s, 3p, and 3d cutoffs of 1.80, 1.80, and 1.80 a.u., respectively, while the oxygen potential has a He core with 2s, 2p, and 3d cutoffs of 1.45, 1.55, and 1.40 a.u., respectively. The pseudopotentials were tested against all-electron DFT computations.

**DFT Computations.** Static DFT computations were performed within the ABINIT (31) package. Computations used the WC with a plane wave energy cutoff of 100 Ha and converged, (0.5, 0.5, 0.5)-shifted Monkhorst-Pack k-point meshes of 4x4x4 for quartz, 4x4x6 for stishovite/$CaCl_2$, and 4x4x4 for $\alpha PbO_2$ primitive cells. The crystal geometries were allowed to relax until all forces were smaller than $10^{-4}$ Hartree/Bohr. Energies were typically computed at six different volumes ±10% about the equilibrium volume to generate the equation of state.

**Phonon Computations.** Phonon free energies were computed with the DFT(WC) linear response method within the quasiharmonic approximation. The quasiharmonic phonons depend on volume but not temperature, and they give low order anharmonic, temperature-dependent contributions to the free energy. A plane-wave cutoff energy of 40 Ha with 4x4x4 q-point and k-point meshes were found to give converged results.

**QMC computations.** Quantum Monte Carlo computations were performed using the CASINO code (32). We used the pseudopotentials constructed for the DFT(WC) computations and the crystal geometries used in QMC are fixed to be the relaxed structures obtained within DFT(WC). At the start of a QMC computation, a trial wave function is constructed from the Slater determinant of single particle orbitals from the corresponding DFT(WC) computations. The QMC trial wave function is constructed by multiplying the determinant of DFT orbitals, each represented in a B-spline basis, by a Jastrow correlation factor containing two-body (e-e, e-n), three-body (e-e-n) and plane-wave expansion terms. The Jastrow factor contains 44 optimizable parameters.

The parameters in the Jastrow factor were optimized by minimizing the variance of the VMC energy in a cyclic procedure which is terminated when the fluctuations in the total energy are smaller than a two sigma statistical error. DMC allows the wave function to evolve according to the imaginary-time Schrödinger equation, projecting out the ground state. The only essential approximation in DMC is the so-called fixed node approximation, in which the nodes of the wave function are fixed to equal those of the trial (Slater determinant of DFT orbitals) wave function.

Finite size errors were reduced below 30 meV by using simulation cells up to 72 atoms (2x2x2) for quartz, 162 atoms (3x3x3) for stishovite/$CaCl_2$, and 96 atoms (2x2x2) for $\alpha PbO_2$, along with using the model periodic Coulomb Hamiltonian (33). A small DMC time step of 0.003 a.u. converges the energy bias below 30 meV. The non-local pseudopotential energy was calculated using the semi-localization scheme of Casula (34). A correction for errors in k-point sampling was estimated from the difference between DFT total energies of a converged k-mesh and a mesh corresponding to the QMC calculation. Finally, a kinetic energy correction due to long range correlations was added to the QMC energy.

**ACKNOWLEDGEMENTS.** This research is supported by the NSF (EAR-0530282, EAR-0310139) and the DOE (DE-FG02-99ER45795). A portion of the computational resources were provided through the "Breakthrough Science Program" of the Computational and Information Systems Laboratory of the National Center for Atmospheric Research which is supported by the National Science Foundation. Significant resources were also provided through "friendly user" alpha- and beta-testing and production time at NERSC. Additional computational resources were provided by TeraGrid, NCSA, OSC, and CCNI. We acknowledge Ken Esler, Richard Hennig and Neil Drummond for helpful discussions.

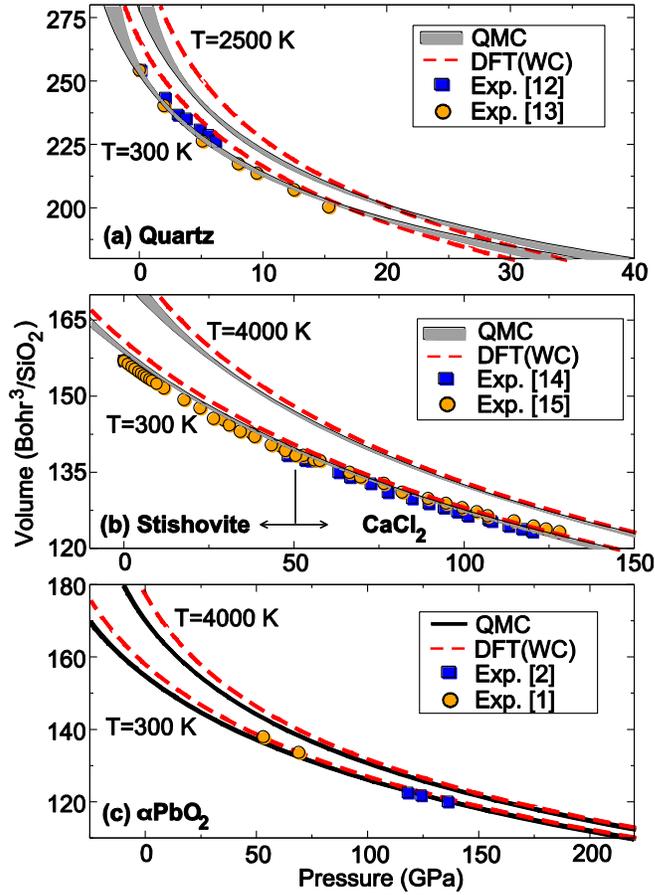

FIG. 1: Thermal equations of state of (a) quartz, (b) stishovite and $CaCl_2$, and (c) $\alpha PbO_2$. The lower sets curves in each plot are room temperature and the upper sets are near the melting temperature. Gray shaded curves are QMC results, with the shading indicating one-sigma statistical errors. The dashed lines are DFT results using the WC functional. Symbols represent diamond-anvil-cell measurements (Exp.) (1, 2, 12-15).

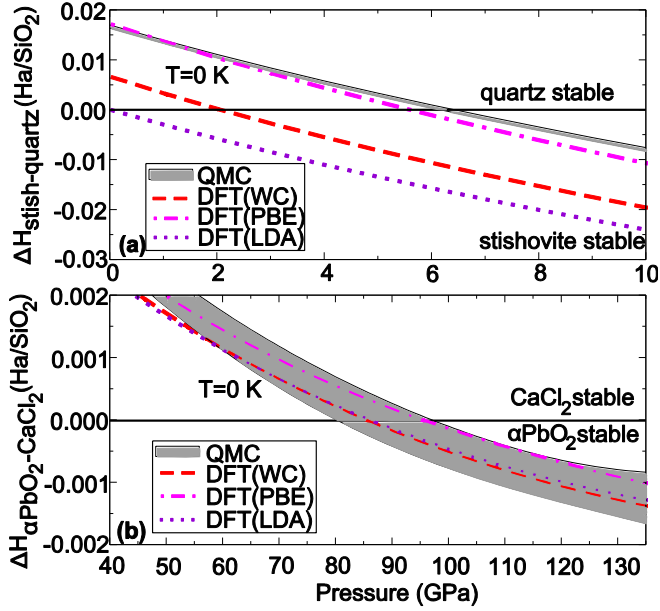

FIG. 2: Enthalpy difference of the (a) quartz-stishovite, and (b) $CaCl_2$-$\alpha PbO_2$ transitions. Gray shaded curves are QMC results, with the shading indicating one-sigma statistical errors. The dashed, dot-dashed, and dotted lines are DFT results using the WC, PBE, and LDA functionals, respectively.

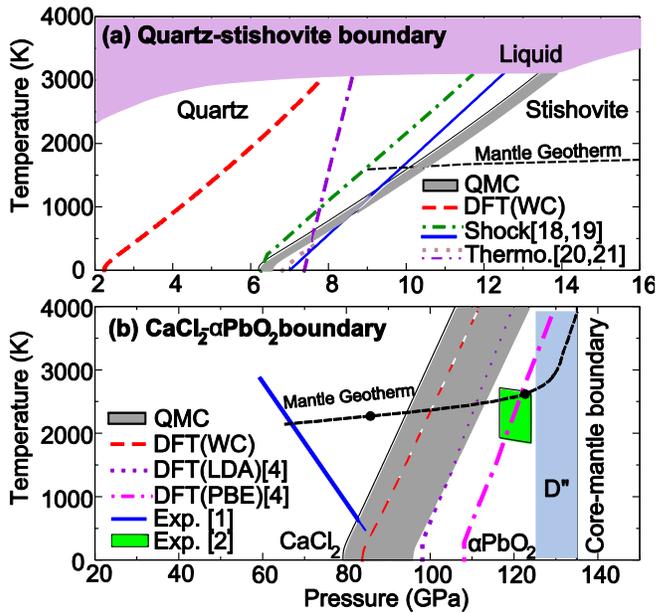

FIG. 3: (a) Computed phase boundary of the quartz-stishovite transition. The gray shaded curve is the QMC result, with the shading indicating one-sigma statistical errors. The dashed line is the boundary predicted using WC. The dash-dot and solid lines represent shock (19, 20) analysis, while dotted and dash-dash-dot lines represent thermochemical data (Thermo.) (21, 22). QMC agrees well with the shock and thermochemical boundaries, while the WC boundary is about 4 GPa too low. The QMC boundary also crosses the melting curve (23) at the quartz-stishovite melt intersection. (b) Computed phase boundary of the $CaCl_2$-$\alpha PbO_2$ transition. Gray shaded curves are QMC results, with the shading indicating one-sigma statistical errors. The dashed, dotted, and dash-dot lines are DFT boundaries using WC, LDA (4), and PBE (4) functionals, respectively. The dark green shaded region and the solid blue line are diamond-anvil-cell measurements (Exp.) (1, 2), while the vertical light blue bar represents pressures in the the D″ region. The measurements spread from 65 to 120 GPa at the mantle geotherm. QMC constrains the transition at the mantle geotherm to 105(8) GPa. Circles drawn on the geotherm indicate the two-sigma QMC boundary.

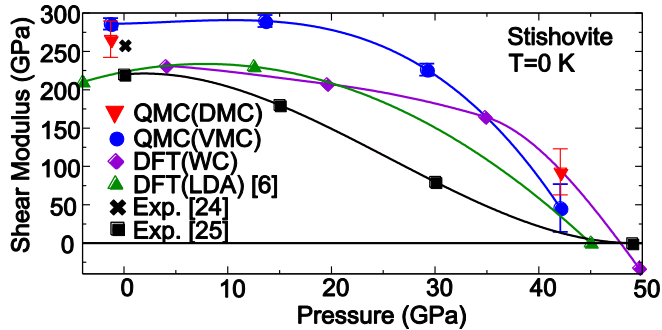

FIG. 4: Softening of the shear modulus of stishovite with pressure. Down triangles and circles represent QMC results using the DMC and VMC techniques, respectively. Diamonds and up triangles represent DFT results within the WC and LDA (6). Squares show radial X-ray diffraction data (24) and crosses at zero pressure show Brillouin scattering data (25). The shear modulus in all methods softens rapidly with increasing pressure and becomes unstable near 50 GPa, signaling a transition to the $CaCl_2$ phase.

**TABLE 1. Computed QMC thermal equation of state parameters at ambient conditions (300 K, 0 GPa).**

| Phase | quartz | | stishovite | | $\alpha PbO_2$ | |
|---|---|---|---|---|---|---|
| Method | QMC | Exp. | QMC | Exp. | QMC | Exp. |
| $F_0$ | -35.7946(2) | | -35.7764(3) | | -35.7689(2) | |
| $V_0$ | 254(2) | [a]254.32 | 159.0(4) | [d,e]157.3-156.94 | 154.8(1) | [h]157.79 |
| $K_0$ | 32(6) | [a]34 | 305(20) | [d,e]291-309.9(1.1) | 329(4) | [h]313(5) |
| $K'_0$ | 7(1) | [a]5.7(9) | 3.7(6) | [d,e]4.29-4.59 | 4.1(1) | [h]3.43 (11) |
| $\alpha \times 10^{-5}$ | 3.6(1) | [b]3.5 | 1.2(1) | [f]1.26(11) | 1.2(1) | |
| $C_p/R$ | 1.82(1) | [b]1.80 | 1.71(1) | [f]1.57(38) | 1.69(1) | |
| $\gamma$ | 0.57(1) | [c]0.57 | 1.22(1) | [f,g]1.35-1.33(6) | 1.27(1) | |
| $q$ | 0.40(1) | [c]0.47 | 2.22(1) | [f,g]2.6(2)-6.1 | 2.05(1) | |
| $\delta_T$ | 6.27(1) | [b]3.3-8.9 | 5.98(1) | [f,g]6.6-8.0(5) | 6.40(1) | |

a. Reference 13

b. Reference 26

c. Reference 27

d. Reference 14

e. Reference 15

f. Reference 28

g. Reference 29

h. Reference 1